\documentclass[preprint2,longabstract]{aastex}

\slugcomment{Not to appear in Nonlearned J., 45.}

\shorttitle{Distance and Proper Motion Measurement of S~Per}
\shortauthors{Asaki et al.}

\begin{document}


\title{
  Distance and Proper Motion Measurement \\
  of the Red Supergiant, S~Persei, \\
  with VLBI H$_{2}$O Maser Astrometry 
}


\author{Y. Asaki\altaffilmark{1,2}}
\affil{Institute of Space and Astronautical Science, 
       3-1-1 Yoshinodai, Chuou, Sagamihara, Kanagawa 252-5210, Japan}
\email{asaki@vsop.isas.jaxa.jp}

\author{S. Deguchi\altaffilmark{3}}
\affil{Nobeyama Radio Observatory, 
       Nobeyama, Minamimaki, Minamisaku 384-1305, Japan}
\email{deguchi@nro.nao.ac.jp}

\author{H. Imai\altaffilmark{4}}
\affil{
       Department of Physics and Astronomy, 
       Graduate School of Science and Engineering,
       Kagoshima University, 
       1-21-35 Korimoto, Kagoshima 890-0065, Japan}
\email{hiroimai@sci.kagoshima-u.ac.jp}

\author{K. Hachisuka\altaffilmark{5}}
\affil{Shanghai Astronomical Observatory,
       Chinese Academy of Sciences, 
       Shanghai 200030, China}
\email{khachi@shao.ac.cn}

\author{M. Miyoshi\altaffilmark{6}}
\affil{Division of Radio Astronomy, 
       National Astronomical Observatory of Japan, 
       2-21-1 Osawa, Mitaka, Tokyo 181-8588, Japan}
\email{makoto.miyoshi@nao.ac.jp}

\and

\author{M. Honma\altaffilmark{7}}
\affil{Mizusawa VLBI Observatory,
       National Astronomical Observatory of Japan, 
       2-21-1 Osawa, Mitaka, Tokyo 181-8588, Japan}
\email{mareki.honma@nao.ac.jp}


\altaffiltext{1}{
    Institute of Space and Astronautical Science, 
    3-1-1 Yoshinodai, Chuou, Sagamihara, Kanagawa 252-5210, Japan
}
\altaffiltext{2}{
    Department of Space and Astronautical Science,
    School of Physical Sciences, 
    The Graduate University for Advanced Studies (SOKENDAI),
    3-1-1 Yoshinodai, Chuou, Sagamihara, Kanagawa 252-5210, Japan
}
\altaffiltext{3}{
    Nobeyama Radio Observatory, 
    Nobeyama, Minamimaki, Minamisaku 384-1305, Japan
}
\altaffiltext{4}{
    Department of Physics and Astronomy, 
    Graduate School of Science and Engineering,
    Kagoshima University, 
    1-21-35 Korimoto, Kagoshima 890-0065, Japan
}
\altaffiltext{5}{
    Shanghai Astronomical Observatory,
    Chinese Academy of Sciences, 
    Shanghai 200030, China
}
\altaffiltext{6}{
    Division of Radio Astronomy, 
    National Astronomical Observatory of Japan, 
    2-21-1 Osawa, Mitaka, Tokyo 181-8588, Japan
}
\altaffiltext{7}{
    Mizusawa VLBI Observatory,
    National Astronomical Observatory of Japan, 
    2-21-1 Osawa, Mitaka, Tokyo 181-8588, Japan
}


\begin{abstract}
We have conducted 
VLBA 
phase-referencing monitoring of H$_{2}$O masers
around the red supergiant, S~Persei, for six years.
We have fitted maser motions to a simple expanding-shell
model with a common annual parallax and stellar proper motion,
and obtained the annual parallax
as $0.413 \pm 0.017$~mas, and the stellar proper motion as
($-0.49 \pm 0.23$~mas~yr$^{-1}$,~$-1.19 \pm 0.20$~mas~yr$^{-1}$)  
in right ascension and declination, 
respectively. 
The obtained annual parallax corresponds to the trigonometric
distance of 
$2.42^{+0.11}_{-0.09}$~kpc.  
Assuming the Galactocentric distance of the Sun of 8.5~kpc,
the circular rotational velocity of the LSR at the 
distance of the 
Sun of 220~km~s$^{-1}$,
and a flat Galactic rotation curve,
S~Persei is suggested to have a non-circular motion 
deviating from the Galactic circular rotation for 15~km~s$^{-1}$, 
which is mainly dominated by the anti rotation direction component 
of $12.9 \pm 2.9$~km~s$^{-1}$. 
This red supergiant is thought 
to belong to the OB association, Per~OB1, so that this non-circular 
motion is representative of a motion of the OB~association in the 
Milky Way. 
This non-circular motion is 
somewhat 
larger than that explained by the standard density-wave theory 
for a spiral galaxy, and
is attributed to either a cluster shuffling of the OB association, or to 
non-linear interactions between non-stationary spiral arms and multi-phase 
interstellar media. 
The latter 
comes from a new view of a spiral arm formation in the 
Milky Way 
suggested
by 
recent large N-body/smoothed particle hydrodynamics
numerical simulations.
\end{abstract}





\section{
  Introduction
}\label{sec:01}

Phase-referencing VLBI technique is capable of 
determining positions of celestial objects 
relative to background calibrators with sub-milliarcsecond accuracy 
\citep{Bartel1986}. 
Absolute position changes of Galactic maser sources can be directly 
detected on the sky using this technique, 
and the observables are very useful for measuring annual parallaxes 
or trigonometric distances to the sources 
and their proper motions 
\citep[e.g.,][]{Xu2006, Hachisuka2006}.  
In addition, the measured trigonometric distance and proper motion 
together with the radial velocity of a source can provide us 
a complete three-dimensional kinematic information without 
model 
assumptions. 

\begin{figure}
\epsscale{.45}
\plotone{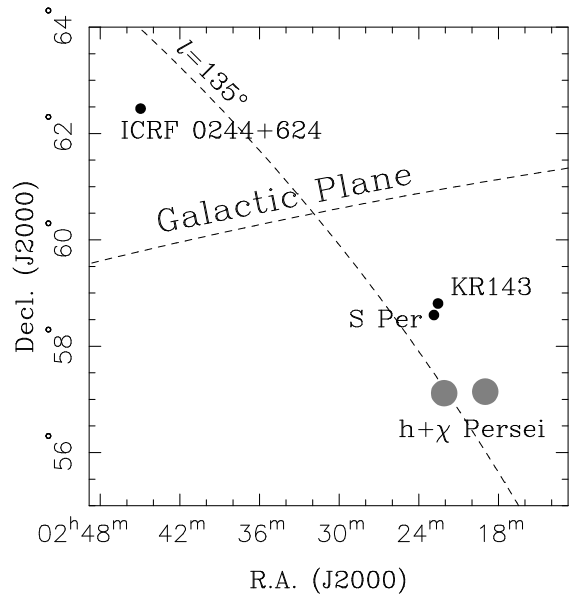}
\caption{
    Sky position of the observed sources in the present VLBI 
    phase-referencing monitoring. The position of $h+\chi$~Per 
    is also indicated. 
}
\label{fig:01}
\end{figure}

There are a number of Galactic evolved stars with circumstellar 
H$_2$O masers, which are observable with VLBI at 22~GHz. 
The red supergiant, S~Persei 
(hereafter, S~Per), 
is one such target harboring 
H$_2$O masers as well as SiO and OH masers 
\citep[e.g.,][]{Alcolea1990, Richards1996}. 
This star is separated by $1.5^{\circ}$ from a double cluster, 
$h$ and $\chi$~Persei (hereafter, $h+\chi$~Per, see Figure~\ref{fig:01}), 
located at the outer edge of the Perseus arm, 
which involves a  large group of more than 20 red supergiant stars. 
The cluster is the most massive open cluster within 3~kpc from 
the Sun and belongs to the OB association, Per~OB1. 
\citet{Humphreys1975} 
reported that the radial velocity of S~Per 
is fairly consistent with 
that of $h+\chi$~Per, so that S~Per is 
thought to belong to the OB association. We can therefore determine 
the distance and three-dimensional motion of the OB association 
in the spiral arm by measuring the annual parallax and proper motion 
of the member star. 

We report 
results of 
phase-referencing VLBI monitoring of the H$_2$O masers 
associated with 
S~Per 
over 
six years with the 
Very Long Baseline Array (VLBA) of the 
National Radio Astronomy Observatory (NRAO). 
The observations are described in Section~\ref{sec:02}. 
Data reduction including astrometric analyses is 
presented 
in Section~\ref{sec:03}. 
The results are presented in Section~\ref{sec:04}. We discuss the results 
in Section~\ref{sec:05} and summarize this study in Section~\ref{sec:06}.  
The results 
presented here replace those 
reported previously 
by 
\citet{Asaki2007a}. 
In this paper, we define that the Local Standard of Rest (LSR) is 
a reference frame at the position of the Sun and moving in a circle 
around the center of the 
Milky Way. 
We adopt the IAU standard values for the Galactocentric distance 
of the Sun, 
$R_0=8.5$~kpc, 
and the circular rotational velocity of the LSR at the 
distance of the 
Sun, 
$\Theta_0=220$~km~s$^{-1}$. 
We also assume 
a flat Galactic rotation curve around the Sun. 
For almost all the parts, we adopt a Solar motion of 
20~km~s$^{-1}$ relative to the LSR to the direction of 
($\alpha_{\mathrm{1900}}$,~$\delta_{\mathrm{1900}}$)= 
($18^{\mathrm{h}}$,~$30^{\circ}$) for the radial velocity 
(radio definition of the LSR). 
Hereafter this Solar motion is referred to as the standard 
Solar motion, and the LSR with the standard 
Solar motion is referred to as the traditional 
LSR. In the following analyses, 
we adopt the systemic velocity of S~Per 
of $-38.5$~km~s$^{-1}$ 
\citep{Diamond1987} 
in the traditional LSR. On the other hand, 
%
%
because the Galactic parameters are known to involve some uncertainties 
\citep{McMillan2010}, 
we discuss the observational results adopting more realistic values 
for the Solar motion in Section~\ref{sec:05-02}, 
which was determined from the average motion of a number of stars 
in the solar neighborhood observed with the Hipparcos satellite. 
For the proper motion, we adopt the heliocentric 
coordinates. 

\section{
  Observations
}\label{sec:02}

Figure~\ref{fig:01} shows sky positions of the observed sources. 
VLBI phase-referencing observations 
of S~Per 
at 22.2~GHz 
have been conducted together with a closely located continuum source 
as a positional reference 
at 
a separation angle of 
$0.2^{\circ}$. 
This positional reference source was initially selected from 
the 6-cm northern sky catalogue 
\citep{Becker1991} 
and the Texas survey 
\citep{Douglas1996}. 
Later we found that this source is catalogued 
under the number 143 
in 
the 
21-cm radio continuum survey of the Galactic plane by 
\citet{Kallas1980} 
(the so-called ``KR survey'') 
and an outer Galaxy VLA survey by 
\citet{Fich1986} 
based on the KR survey, so that the source is referred to hereafter as KR143. 
\cite{Imai2001} 
for the first time confirmed that KR143 
is detectable with VLBI observations at 22~GHz, while we could 
not find an optical counterpart for KR143. 
We have conducted a total of eight-epoch VLBI observations over 
six years with the ten VLBA antennas. 
The 
observing 
epochs are listed in Table~\ref{tbl:01}. 

Two intermediate-frequency (IF) bands with 16-MHz bandwidth 
in left-hand circular polarization were recorded. 
These 
IF bands 
are continuous in frequency. The H$_2$O maser emissions of S~Per 
has been received with the lower IF band 
(IF1) while the upper IF band 
(IF2) was used to improve the Signal-to-Noise Ratio 
(SNR) for KR143. We have also conducted phase-referencing observations 
of a pair of KR143 and ICRF~0244+624 with a separation angle of 
$4.8^{\circ}$ in the last five epochs to check the position stability 
of KR143 with respect to the positionally reliable ICRF source. 
A larger separation angle will take a longer telescope slew time, 
and therefore a longer switching 
duty cycle.  
In general, the larger the separation angle 
and/or the longer the switching 
duty cycle 
is, the more difficult 
is phase-referencing 
at 22~GHz due to stochastic tropospheric 
phase fluctuations. We selected a lower observing frequency of 15.3~GHz 
for the pair of KR143 and ICRF~0244+624 to 
reduce 
a coherence loss 
effect due to the tropospheric phase fluctuations. 
The switching cycle time, 
$t_{\mathrm{s}}$, 
for the phase-referencing is listed in Table~\ref{tbl:01}. 

In the first three epochs, 20-min phase-referencing sequence at 22~GHz 
was interleaved by 2- to 3-min calibrator observations. 
In 
the last five epochs, 20-min phase-referencing sequences 
at 22~GHz and 15~GHz were conducted 
in 
turn, and each sequence was 
interleaved by short calibrator scans at the same frequency. 
VLBI cross-correlation was 
carried out at the VLBA correlation center in Socorro 
to generate VLBI fringe data with 1024-frequency channels, 
corresponding to 
a velocity resolution of 0.2107~km~s$^{-1}$. 
In the first-epoch cross correlation, a 
wrong position with an offset by 4~arcsec 
was 
used 
as a phase tracking center of S~Per. This mistake was 
recovered in 
the course of 
data reduction process. 
The observed pairs of sources in our monitoring program are listed in 
Table \ref{tbl:02} together with their separation angles. 

\section{
  Data reduction
}\label{sec:03}

We analyzed the VLBI 
data 
using the standard NRAO data reduction software, 
Astronomical Image Processing Software (AIPS), version 31DEC06. 
There were two data reduction series: 
one for the pair S~Per and KR143, and 
another one 
for the pair KR143 and ICRF~0244+624. 

\subsection{
  Initial calibrations
}\label{sec:03-01}

Initial calibration tasks were common 
for both 
the pairs of sources. 
First, 
the amplitude calibration was 
conducted by 
using the 
measured system noise temperatures and gain 
variation data. In the next step, 
Earth Orientation Parameters (EOP) errors, two-bit sampling bias in 
the analogue-to-digital (A/D) conversion in VLBI signal processing, 
antenna parallactic angles, and ionospheric dispersive delays were 
corrected. 

In the first three epochs, NRAO~150, J0102+5824 and 3C~84 have been 
observed as calibrators at 22.2~GHz, while 
the former two 
have been observed as calibrators in the last five 
epochs. 
At 15.3~GHz, NRAO~150 
and J0102+5824 
have been 
observed as calibrators in the last five epochs. Among them, 
NRAO~150 was used both in the data reduction series for calibrating 
relative clock offsets and phase offsets between the independent antennas. 
Phase differences between the two IF bands were also removed by 
using the 
NRAO~150 fringe data. 
In the 
next steps of 
data analysis, a customized data reduction path was used for each 
pair of the sources. 

\subsection{
  Data reduction for S~Per and KR143 at 22~GHz
}\label{sec:03-02}

For the maser data analysis, the Doppler shift in S~Per spectra 
due to the Earth rotation and Earth's Kepler motion were corrected with 
the AIPS spectrum channel correction task, CVEL. The observed frequencies 
of the maser lines were converted to radial velocities with respect 
to the traditional LSR using 
the 
rest frequency of 22.235080~GHz for the 
H$_2$O $6_{16}-5_{23}$ transition. The spectrum deformation due to the 
receiver complex gain characteristics was calibrated with 
the NRAO~150 spectrum in the CVEL process. 

The 4 arcsec error in the position of 
S~Per's phase tracking center 
in the first epoch was corrected with the AIPS visibility 
phase correction task UVFIX. Note that, 
although the S~Per visibilities might be subject to the coherence 
loss 
as a result of the cross correlation task with the phase tracking 
center error, the following astrometric analysis is not affected. 

We inspected the cross-power spectra for all the baselines 
in order to select frequency channels of emission which 
would be unresolved and not have rapid amplitude--time variation. 
We then carried out fringe-fitting 
by using the AIPS task FRING, with an option of no delay search for the 
selected channel. The fringe-fitted data were image-synthesized with a 
50-$\mu$as pixel for a region of $4096\times4096$ pixels, roughly 
a $200\times200$-square mas region by using the AIPS task IMAGR. 
Using this image as an initial model, phase and amplitude 
self-calibration was conducted. 

With the above processes from the fringe-fitting to the self-calibration, 
we obtained the complex gain calibration data, which were applied to the 
remaining frequency channels to make image cubes for the same maser region. 
The images were 
produced for all channels separately 
in the velocity range of $-23$ and $-61$~km~s$^{-1}$ 
with the step of 0.2107~km~s$^{-1}$ 
in the traditional LSR. To pick up maser emission initially, 
we used the 2-D Gaussian component survey task SAD in each 
velocity channel map. Based on the results of SAD, we visually inspected 
the surveyed Gaussian components one by one. 
We stored the image pixels larger than 3-$\sigma$ noise level for a 
visually confirmed maser emission. Hereafter we define a maser 
``spot'' as an emission in a single velocity channel, and 
a maser ``feature'' as a group of spots 
observed in at least two continuous velocity channels at a coincident 
or very closely located 
positions. 

For 
KR143 imaging, 
the Doppler correction and bandpass calibration were skipped because 
the spectrum was averaged for the continuum 
source. Since there was no S~Per's maser emission in IF2, 
the complex gain calibration data for IF1 was applied to IF2, 
whose phase and delay biases from IF1 was already removed in the initial 
calibration. 
The synthesized image of KR143 with the pixel size of 50~$\mu$as was 
obtained by 
using 
two IF bands, 
so that the image SNR is higher than that with the single IF band by 
a factor of 
$\sqrt{2}$. 
The SNR of KR143 phase-referenced images ranges between 18 and 37 
at 22~GHz. 

\subsection{
  Data reduction for KR143 and ICRF~0244+624 at 15~GHz
}\label{sec:03-03}

The Doppler and bandpass correction were skipped for both sources 
in processing of the 15~GHz data as unnecessary. 
ICRF~0244+624 was imaged with the 50-$\mu$as pixel size 
to obtain the complex gain calibration data. This calibration 
data were applied to the KR143 fringe at 15~GHz before imaging. 
Synthesized images of KR143 were made with the 50-$\mu$as pixel size. 
The SNR of KR143 phase-referenced images ranges between 7 and 17 
at 15~GHz. 

\section{
  Results
}\label{sec:04}

\subsection{
  KR143's position
}\label{sec:04-01}

Phase-referenced radio maps of KR143 are dominated by a weak single 
component both at 15 and 22~GHz. For the present astrometric study, 
we firstly determined the 
position of KR143 with respect to ICRF~0244+628 at 15~GHz, which 
had not been determined to high precision 
previously. 
In this paper, the position of ICRF~0244+624 in J2000 is 
adopted 
as 
$2^{\mathrm{h}} 44^{\mathrm{m}} 57^{\mathrm{s}}.696746$ 
and 
$+62^{\circ} 28' 6''.515030$ 
in right ascension and declination, respectively. 
Since ICRF~0244+624 has shown a very simple structure in the radio maps, 
we treated the peak position of the image as a positionally stable point 
in the sky. 
We assume that KR143's position at 22~GHz is consistent with that 
at 15~GHz, and astrometric analysis of S~Per was determined with 
respect to this position. 

Since images of KR143 have rather low SNR, a Gaussian fitting for 
the detected component may be affected by a rather severe 
image distortion due to a 
high 
image noise even if the object 
consists of only a single component. In addition, the phase-referencing 
image may be distorted 
by the residual phase errors, introduced, e.g., by  
the static tropospheric delays. 
We used the peak position of KR143's image as the positional reference 
in order to 
minimize 
those effects in determining the position. 
The position 
measurement 
error was calculated from the beamwidth 
over the SNR at the peak. 

\begin{figure}
\epsscale{.50}
\plotone{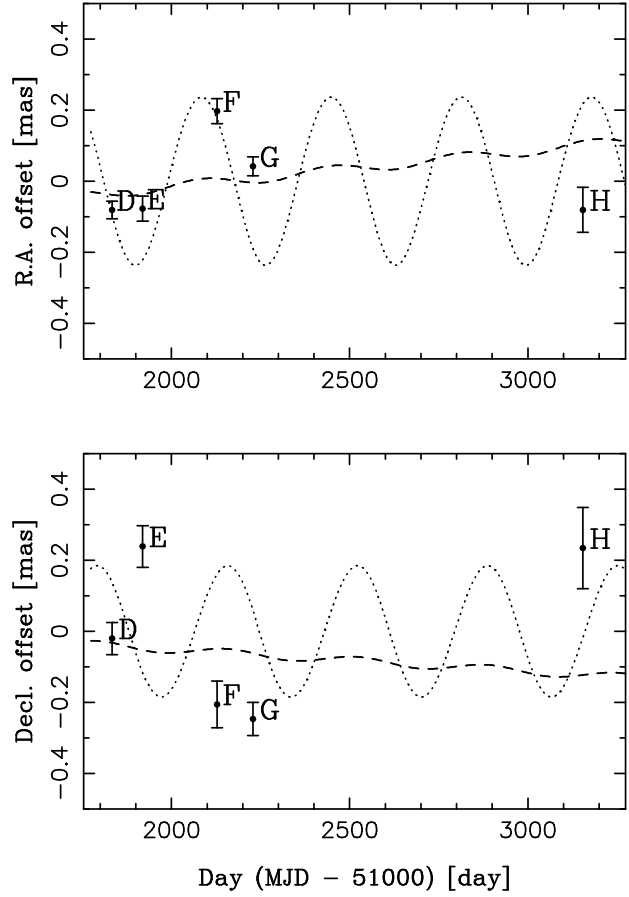}
\caption{
    Peak position of the KR143 phase-referenced images 
    with respect to ICRF~0244+624 at 15~GHz. 
    The top and the bottom plots show the time variation in 
    right ascension and declination, respectively. 
    The zero position in each plot represents the unweighted average 
    for the five epochs. The error bars represent the 
    position measurement error. The dashed lines represent the 
    least-square 
    analysis result for the 
    proper motion and annual parallax. The dotted lines represent an 
    annual parallax for a source at 4-kpc away from the Sun. 
}
\label{fig:02}
\end{figure}

For the astrometric measurements of Galactic sources, a positional 
reference should be a background extragalactic source such as an 
Active Galactic Nuclei (AGN). However, because we hardly know whether 
KR143 is a suitable background source, we then investigated how 
positionally 
stable KR143 is with respect to ICRF~0244+624. Figure~\ref{fig:02} 
shows a time variation of 
the 
KR143's position in the last five epochs 
of our experiment. 
Assuming that 
KR143 is a 
Galactic 
source 
for which the annual parallax is significant, 
we carried out a Levenberg-Marquardt non-linear 
least-square 
analysis to obtain the annual parallax and proper motion, but could not 
find a reliable solution 
because the $\chi^{2}$ value per degree of freedom was quite large, 
no less than 57. 
This suggests that the annual parallax and proper motion are below 
our threshold because KR143 is sufficiently far from the Sun. 
This threshold is determined by the accuracy of the relative 
position measurements in phase-referencing for the pair of the sources. 
Figure~\ref{fig:02} also displays a sinusoidal pattern of our 
least-square 
analysis result as well as another 
sinusoidal pattern for a source 
at 
the distance of 4~kpc, for comparison. 

\begin{figure}
\epsscale{.35}
\plotone{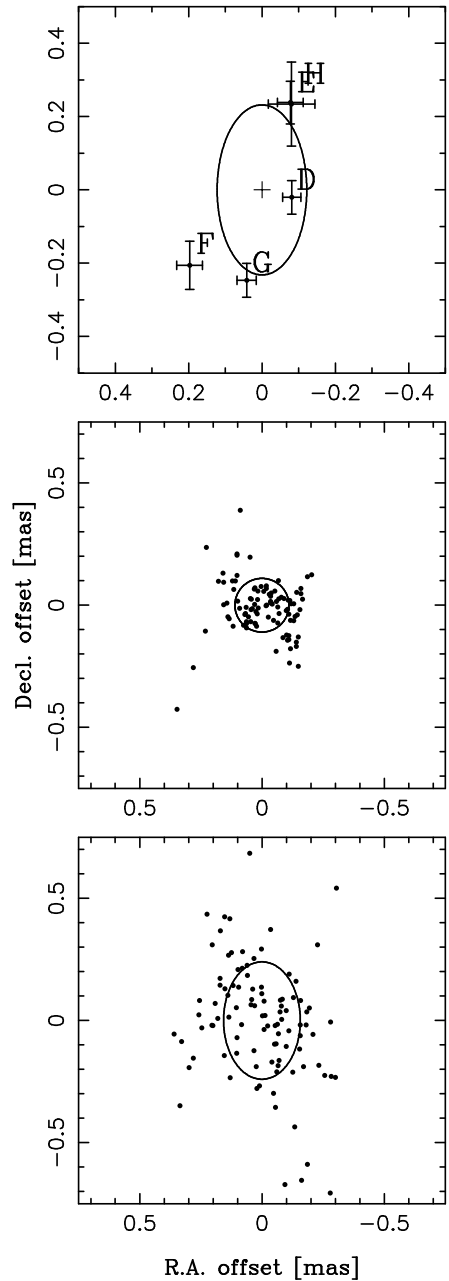}
\caption{
    KR143's image peak position relative to ICRF~0244+624 at 
    15~GHz. The top plot shows the observation results with the error 
    bars, and the middle 
    and bottom plots show results of 100 trials of the observation simulation 
    for KR143 and ICRF~0244+624 with the static tropospheric excess path 
    length error of 3 and 6~cm to the zenith, respectively. The central 
    cross mark at the origin in the top plot is located at 
    the unweighted average position of the measurements. The 
    ellipses represent 1-$\sigma$ for the distributions. 
}
\label{fig:03}
\end{figure}

The spatial distribution of the image 
peaks 
of KR143 at 15~GHz are shown in the top plot of Figure~\ref{fig:03}. 
This plot also shows the 1-$\sigma$ ellipse 
(a solid line).  One possible reason for 
the observed 
change position of the peaks 
is inaccuracies in the VLBI correlator model for the phase-referencing 
astrometry for the pair of the observed sources. 
We conducted Monte Carlo phase-referencing observation simulations 
for the pair of KR143 and ICRF~0244+624 to 
examine 
whether the observed position change 
could 
be explained 
by 
the uncertainties of the VLBI correlator model. 
For generating simulated phase-referencing fringes, 
we used a VLBI observation 
simulator, ARIS 
\citep[Astronomical Radio Interferometer Simulator,][]{Asaki2007b}.  
In the simulations, we assume flux densities of KR143 and 
ICRF~0244+624 of 20 and 500~mJy, respectively, which were determined 
from our observations. 
The 
sources were assumed to be 
point-like 
sources. The 
simulated 
observation schedule and observing system settings such as 
the recording bandwidth and A/D quantization level 
were adopted from actual VLBA observations. 
There were two simulation series: 
one 
included 
1-$\sigma$ static tropospheric excess path length error 
to the zenith (zenith EPL error) of 3~cm 
\citep[e.g.,][]{Reid1999},  
and the other one 
used 
the zenith EPL error of 6~cm. We made 100 trials for each of the zenith 
EPL error cases. The simulation results are shown in the middle 
and bottom plots of Figure~\ref{fig:03} for the 1-$\sigma$ 
zenith EPL error of 3 and 6~cm, respectively. 
In the case of a zenith EPL error of 6~cm, the simulation 
results are consistent with the observation results. 
Since 
some of our observations were conducted 
during the summer period, 
it is quite natural that the zenith EPL error would be around 6~cm. 
We conclude that the KR143's position change is mainly 
due 
to the zenith EPL error of several centimeters in the VLBI correlator model. 
Hereafter 
we refer a 
relative 
position 
error 
for 
a pair of point sources 
with the phase-referencing due to uncertainties 
in 
a VLBI correlator model 
for the atmospheric EPL, 
antenna positions, 
EOP, and other phase error factors 
as an astrometric 
error. 

The measured flux densities of KR143 are shown in Figure~\ref{fig:04} 
together with 
the data from the previous work by 
\citet{Douglas1996}, 
\citet{Kallas1980}, 
\citet{Condon1998}, 
\citet{Becker1991}, 
\citet{Fich1986}, and 
\citet{Imai2001}. 
Note that, 
for the measurements of the flux density of KR143 
from 
our radio 
images, 
we conducted 
a 
phase-self-calibration 
process. 
Figure~\ref{fig:04} 
resembles 
a synchrotron power-law spectrum, 
making it 
very likely that KR143 is an AGN. 
Although there is currently no concrete evidence that KR143 is 
not located in or near our Galaxy because of the unknown redshift, 
we assume throughout this paper that KR143 is an AGN. 
The relative position of KR143 on the sky with respect to ICRF~0244+624 
is determined with the standard errors of 61 and 116~$\mu$as in 
right ascension and declination, respectively, from all the five epochs. 
The absolute position of KR143 is listed in Table~\ref{tbl:03}. 

\subsection{
  Astrometry of S~Per
}\label{sec:04-02}

In the phase-referencing astrometric analysis for S~Per, 
we investigated the time variation of the peak positions of the maser 
spots with respect to the peak position of the 22-GHz phase-referenced 
radio image of KR143. 
To estimate 
the astrometric error with the VLBI phase-referencing 
for the pair of S~Per and KR143, 
we conducted Monte Carlo simulations of phase-referencing 
observations for the pair of 
these sources 
at 22.2~GHz with ARIS. 
In the simulations, we 
assumed 
that flux densities of S~Per and KR143 
of 1~Jy and 20~mJy, respectively, and that S~Per was a continuum source. 
The tropospheric zenith EPL error of 6~cm was set as found in the 
simulation described above 
for the pair of KR143 and ICRF~0244+624. 
Due 
to the small separation angle and fast antenna switching, 
the simulations show that the 1-$\sigma$ errors in the relative 
position are 6~$\mu$as and 14~$\mu$as in right ascension and declination, 
respectively. We used these standard deviations to provide the astrometric 
error 
for the pair of the sources in the following analysis. 

\begin{figure}
\epsscale{.80}
\plotone{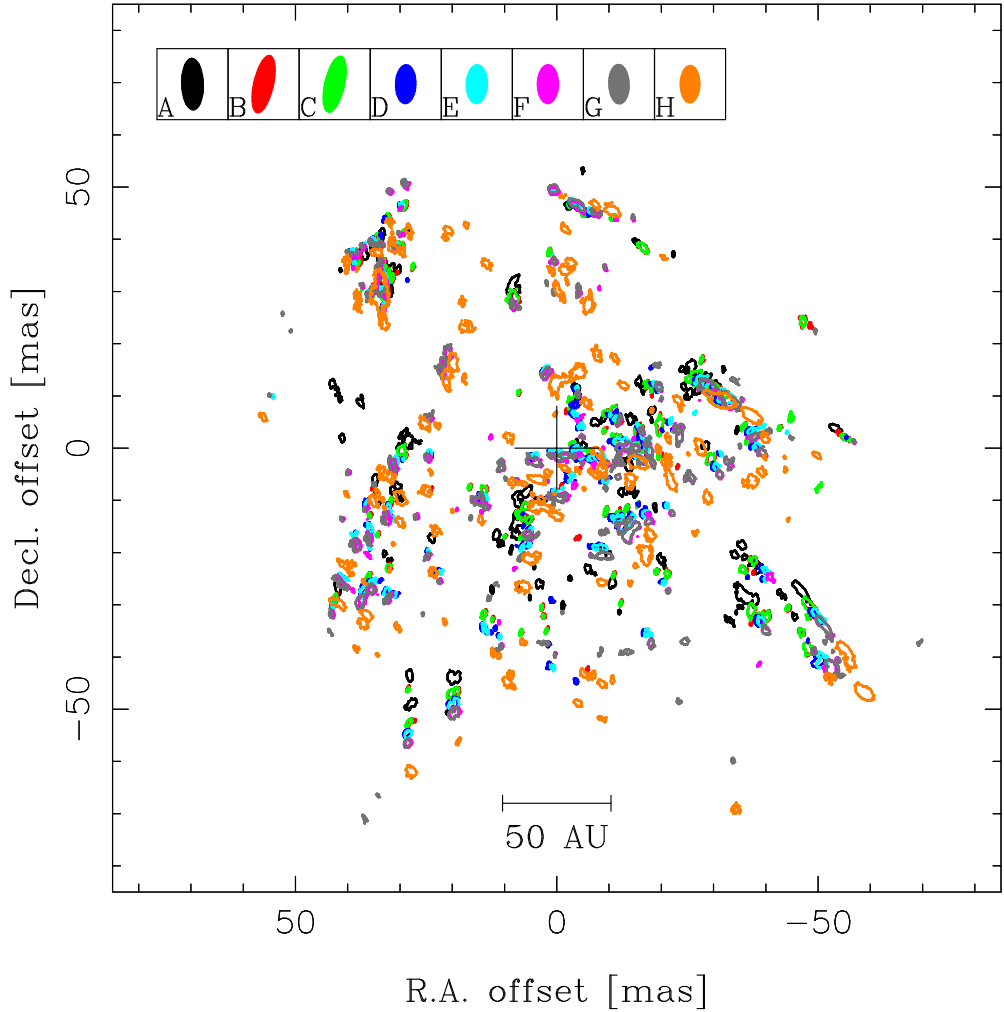}
\caption{
    Spatial distribution of S~Per H$_{2}$O masers in the eight epochs. 
    The central cross represents the error bars of the distribution 
    centroid estimated from the expanding shell flow model fitting 
    described in Section~\ref{sec:04-02}. 
    The outlines of maser features represent a 6-$\sigma$ noise-level 
    contour (90, 138, 156, 150, 168, 150, 162, and 138~mJy 
    at epoch A, B, C, D, E, F, G, and H, respectively). 
    The synthesized beams are shown in the upper 1$\times$1-square mas boxes. 
}
\label{fig:05}
\end{figure}

Figure~\ref{fig:05} shows the spatial distribution of the H$_2$O maser 
emission of S~Per for all the epochs. The maser emission 
is spread 
within a roughly circular structure with a radius of about 
60~mas. We identified 77 maser features consisting of 478 maser 
spots 
in all the available epochs. 
For obtaining the trigonometric parallax 
and proper motion of the star, we conducted the following four steps. 
Firstly, 
we selected a limited number of maser spots to evaluate the annual 
parallax. Our selection criteria are as follows: 
(1) 
maser spots at epochs D, E, F, and G were all detected at the same 
velocity channel with the SNR of at least 6; 
(2) 
in addition to the above epochs, 
they were detected with the SNR of at least 6 at least at one of 
the other epochs at the same velocity channel; 
(3) 
no significant velocity drift larger than 0.25~km~s$^{-1}$~yr$^{-1}$ 
was observed for the features involving the selected spots; and 
(4) 
morphology does not significantly change between the epochs. 
We selected 44 maser spots 
from 
eight maser features. 

\begin{figure}
\epsscale{.80}
\plotone{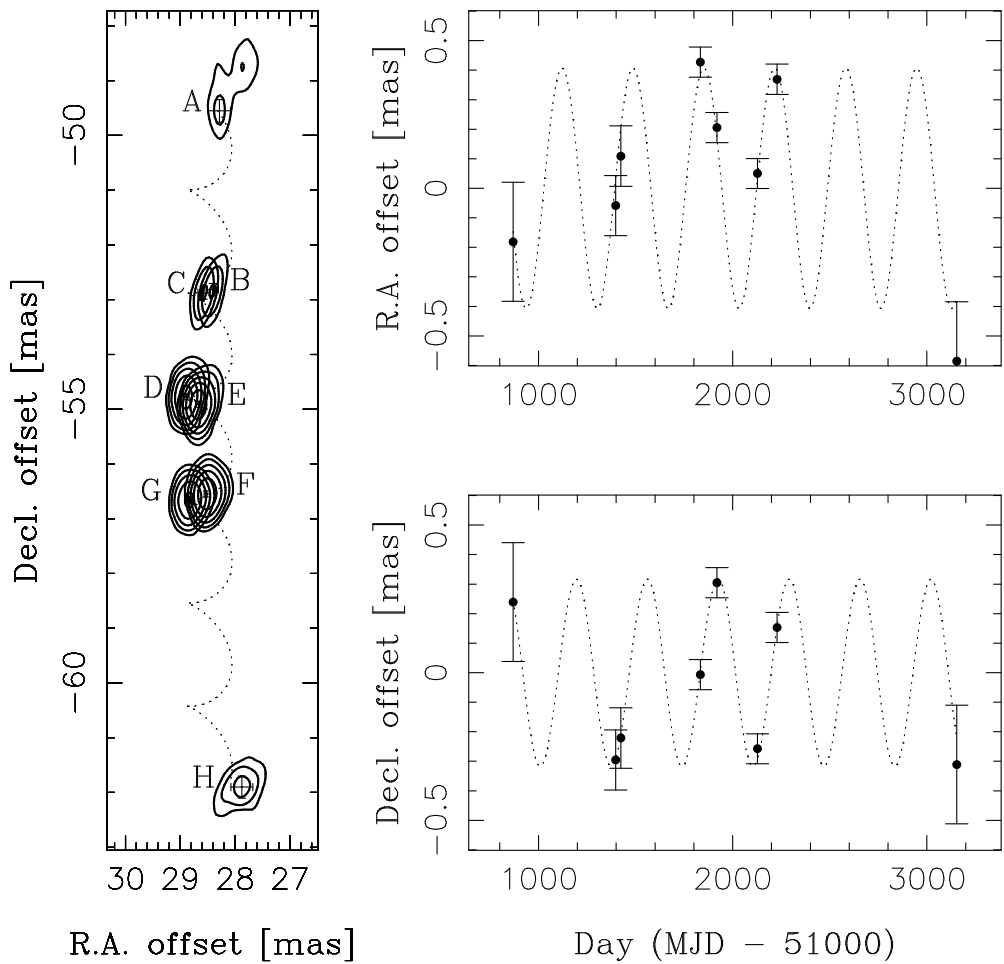}
\caption{
    Left panel: maser spot motion of S~Per on the sky 
    relative to KR143 
    (spot-ID 39 in Table~\ref{tbl:04}). 
    A dotted line represents the best fit annual parallax and 
    proper motion. The outermost contour shows 
    a 6-$\sigma$ noise level for this channel 
    (96, 144, 162, 156, 180, 162, 174, and 168~mJy 
    at epoch~A, B, C, D, E, F, G, and H, respectively), 
    increased by a factor of power of 2. Error bars are 
    root-square-sum of the position measurement 
    error of the brightness peak of the spot, the astrometric error 
    for the pair of the sources, 
    and the diagnostically position error because of 
    the internal structure instability of the maser feature. 
    The right two plots show a residual motion after removing 
    the proper motion. The dotted lines show the best fit annual parallax. 
}
\label{fig:06}
\end{figure}

\begin{figure}
\epsscale{.50}
\plotone{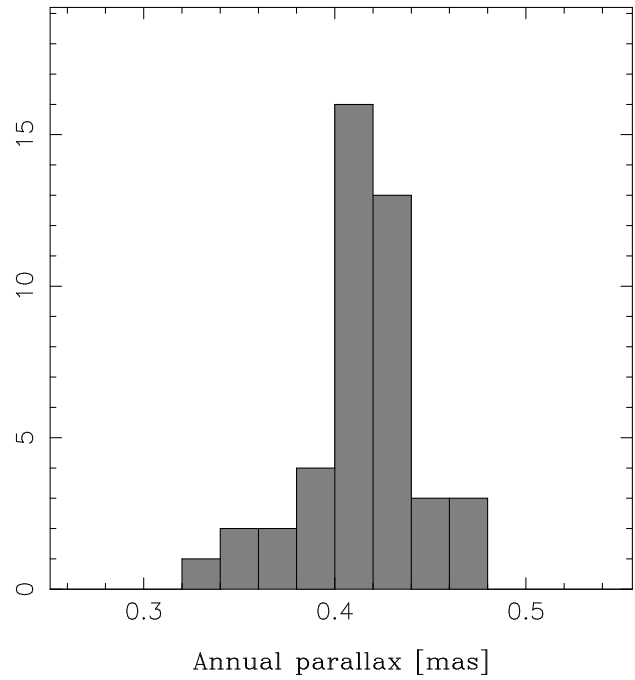}
\caption{
    Histogram of the annual parallaxes for the 44 maser spots listed in 
    Table~\ref{tbl:04}. 
}
\label{fig:07}
\end{figure}

Secondly we carried out the Levenberg-Marquardt 
least-square 
analysis for the i-th maser spot to obtain the initial position, 
($\Delta\alpha^{\mathrm{i}}_{\mathrm{A}}$,~
 $\Delta\delta^{\mathrm{i}}_{\mathrm{A}}$), at epoch A, 
the constant proper motion, 
($\mu^{\mathrm{i}}_{\alpha}\cos{\delta}$,~$\mu^{\mathrm{i}}_{\delta}$), 
along with the annual parallax, $\pi^{\mathrm{i}}$, 
both in right ascension and declination. 
In the initial trial, reduced $\chi^{2}$ values of the analysis 
were in the range from 
a few tens to a few hundreds. 
These large values indicate 
that there must be 
unaccounted 
errors in the analysis. 
As explained above, we expect that the astrometric error in our observation 
is 6 and 14~$\mu$as for right ascension and declination, respectively. 
On the other hand, 
internal morphological time variations of the maser features 
during the monitoring period of over six years cannot be negligible, 
as already pointed out in previous studies
\citep[e.g.,][]{Imai2007}.  
For diagnostic purposes, we 
added position uncertainties of 
0.2~mas for epochs A and H, 
0.1~mas for epochs B and C, and 
0.05~mas for epochs D, E, F, and G to the astrometric 
error 
both for right ascension and declination. 
This 
resulted in making the reduced $\chi^{2}$ values 
equal to unity in our least-square analysis. 
This additional position uncertainty which we attribute to the 
unpredicted time variations in maser morphology is referred to 
hereafter as a maser morphology uncertainty. 
Table~\ref{tbl:04} lists the 
least-square 
analysis results for 
each of the selected maser spots. 
Figure~\ref{fig:06} shows the analysis result of one of the maser spots 
(spot-ID 39 in Table~\ref{tbl:04}). 
Figure~\ref{fig:07} shows the histogram of the evaluated annual parallaxes 
for the 44 maser spots. 
The weighted average of the evaluated annual parallaxes is 
$0.414 \pm 0.006$~mas. 

We thirdly conducted a combined fitting with our 
least-square 
analysis 
for the 44 maser spots to obtain the trigonometric distance to S~Per. 
In this step, we assumed that all the maser spots have a common annual 
parallax, $\pi^{*}$, which is one of the fitting parameters. This 
combined fitting result for $\pi^{*}$ is $0.413$~mas, almost identical 
to 
the weighted average of the 44 annual parallaxes. 
The parameter fitting error in $\pi^{*}$ was $0.006$~mas 
from a covariance matrix of our 
least-square 
analysis. 

\begin{figure}
\epsscale{1.00}
\plotone{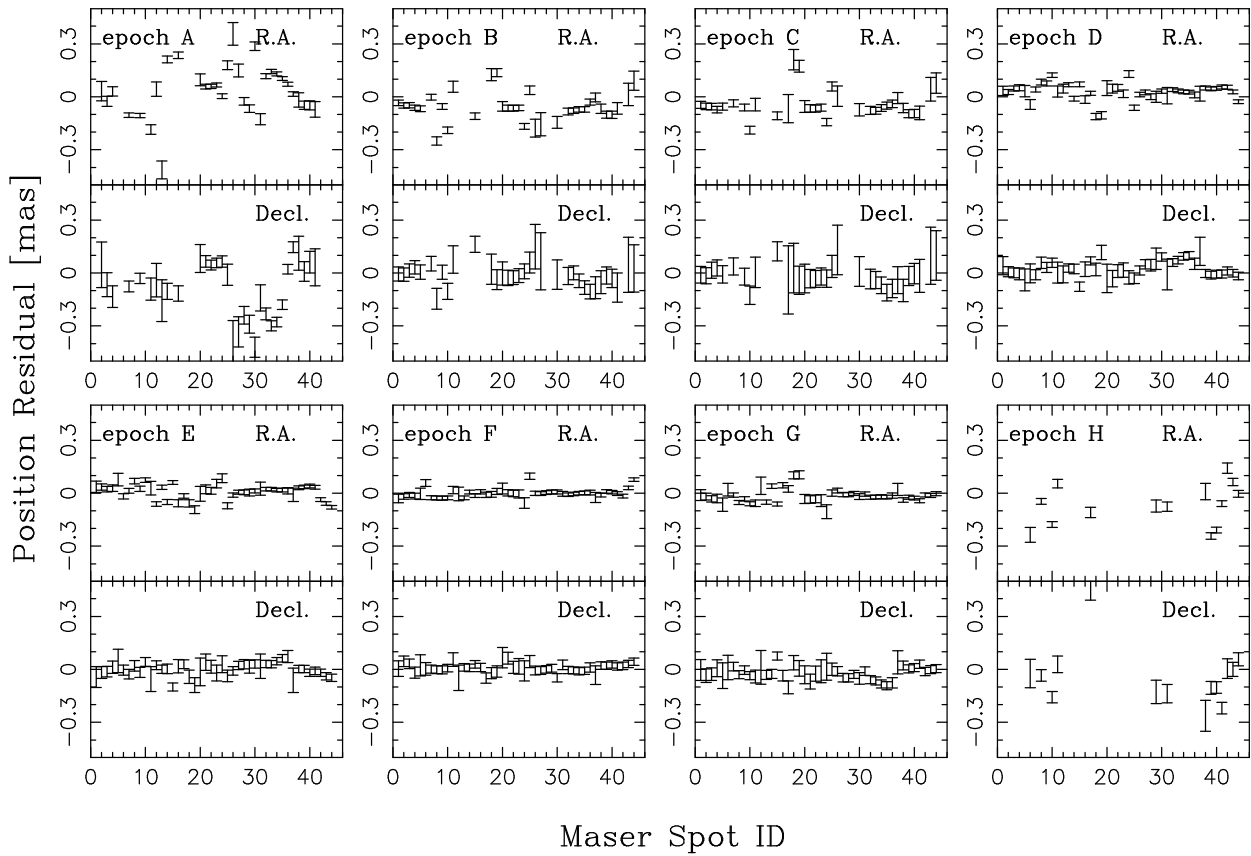}
\caption{
    Residuals in position of the 44 maser spots after 
    removing 
    the obtained annual parallax of 0.413~mas, 
    proper motions, and 
    initial positions. The abscissa represents the maser spot ID 
    as listed in Table~\ref{tbl:04}, and the ordinate is the position 
    residuals in mas. For each epoch, the position residuals 
    in right ascension and declination are shown in 
    the upper and lower part, respectively. Error bars represent the 
    position measurement error. 
}
\label{fig:08}
\end{figure}

In obtaining the annual parallax, two systematic errors have to be 
considered. One is the astrometric error, and the other one is 
the maser morphology uncertainty which provides a common position 
offset to the member 
spots in a certain maser feature. 
The most conservative estimate of the obtained annual parallax error 
becomes 0.040~mas, which is calculated from the above statistical error 
of 0.006~mas multiplying by a factor of $\sqrt{44}$ under an assumption 
that the position error of the selected maser spots 
are not independent. 
\citet{Hachisuka2009} 
applied a method to average positions of a number of maser spots from 
three features of W3(OH) water masers before their parallax fitting 
in order to avoid the maser morphology uncertainty, 
and achieved a 6-$\mu$as accuracy in the annual parallax measurement. 
We note that the impact of the astrometric error on the annual 
parallax measurement is not reduced even if a number of maser features 
are simultaneously processed. On the other hand, the impact of the 
maser morphology uncertainty can be reduced in processing multiple 
maser features because the position offset due to this error 
becomes randomized. Since there are eight maser features in our 
analysis, the error quantity of 0.040~mas is over estimated. 
Figure~\ref{fig:08} shows the position residuals of the 
selected 44 maser spots after removing the obtained annual 
parallax of 0.413~mas, proper motions and initial positions. 
It is noticed that, although a small positive trend can be seen 
in right ascension at epoch D, there are not apparently obvious 
systematic errors at central four epochs from D to G. It is also 
recognizable that systematic errors are patially seen for specific 
maser features possibly due to the time variations in the  maser 
morphology (see, for example, spot-ID 41--44 (feature-ID 8) 
at epochs D, E, and F). 

To evaluate contributions of those systematic errors to the obtained 
annual parallax, we conducted simulations of the combined fitting 
analysis by using fake data sets imitating the selected maser 
spots with the position measurement error, astrometric error, and 
maser morphology uncertainty. A random position offset due to the 
position measurement error to each maser spot at each epoch. 
A position offset due to the astrometric error is applied to all 
the maser spots at each epoch. A position offset due to the maser 
morphology uncertainty is applied to all the maser spots included 
in each of the eight maser features at each epoch. We investigated 
the following error cases: 
(1) only position measurement error; 
(2) only astrometric error; 
(3) only maser morphology uncertainty; and 
(4) a combination of all the three errors above.  
We generated 100 data sets for each of the four cases and carried out 
the Levenberg-Marquardt least-square analysis. 
Root-mean-square errors from the average of the 100 obtained annual 
parallaxes were 3, 9, 15, and 17~$\mu$as for the cases of 1, 2, 3, 
and 4, respectively. The maser morphology uncertainty dominates the 
annual parallax error in our combined fitting while the astrometric 
error corresponds to a relatively small part. It is reasonable to 
adopt the value of 17~$\mu$as as an upper limit of the trigonometric 
distance uncertainty, corresponding to the trigonometric 
distance of $2.42^{+0.11}_{-0.09}$~kpc. We then conducted 
our least-square analysis to obtain individual 
proper motions of all the 478 maser spots with $\pi^{*}$. 

\begin{figure}
\epsscale{.80}
\plotone{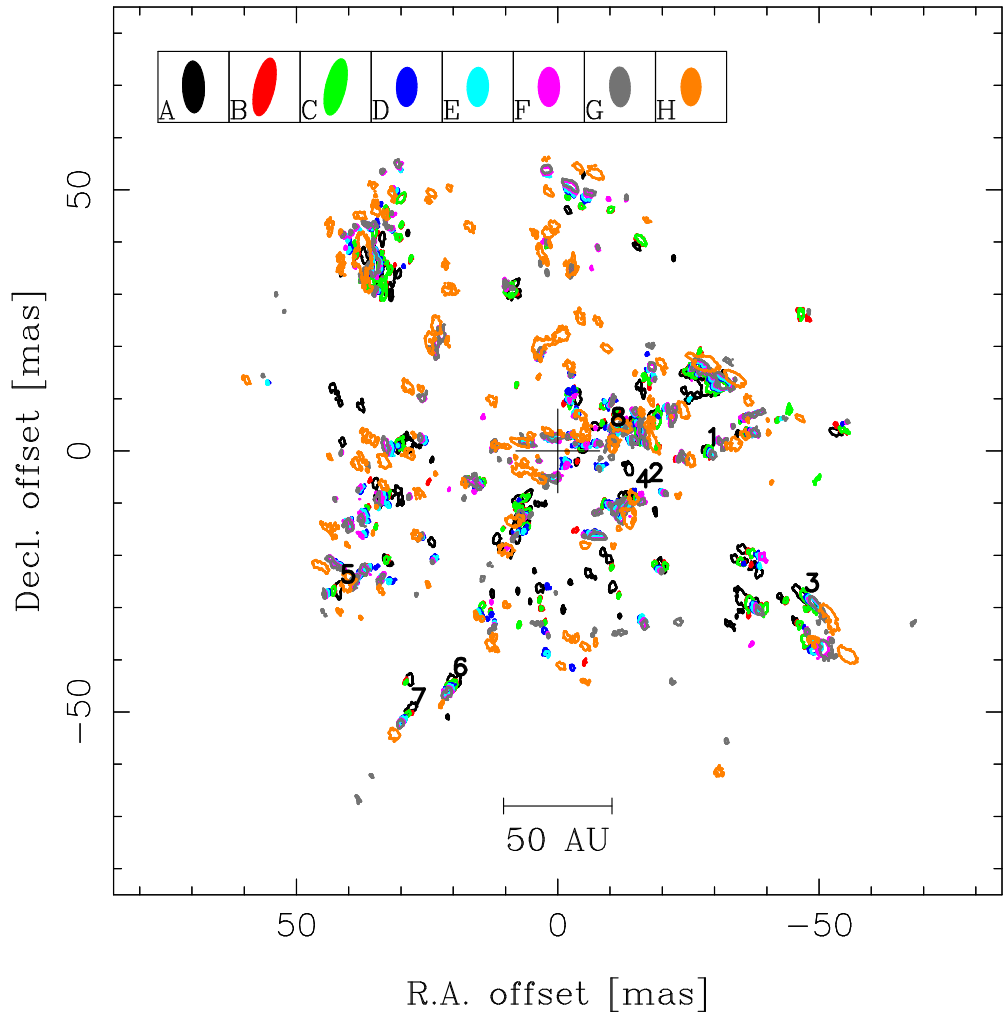}
\caption{
    Same as Figure~\ref{fig:05}, but the annual parallax and 
    stellar proper motion are 
    removed. 
    The numbers in the figure (1 to 8) 
    indicate the maser feature-ID which was used for obtaining the 
    stellar annual parallax. 
}
\label{fig:09}
\end{figure}

\begin{figure}
\epsscale{.80}
\plotone{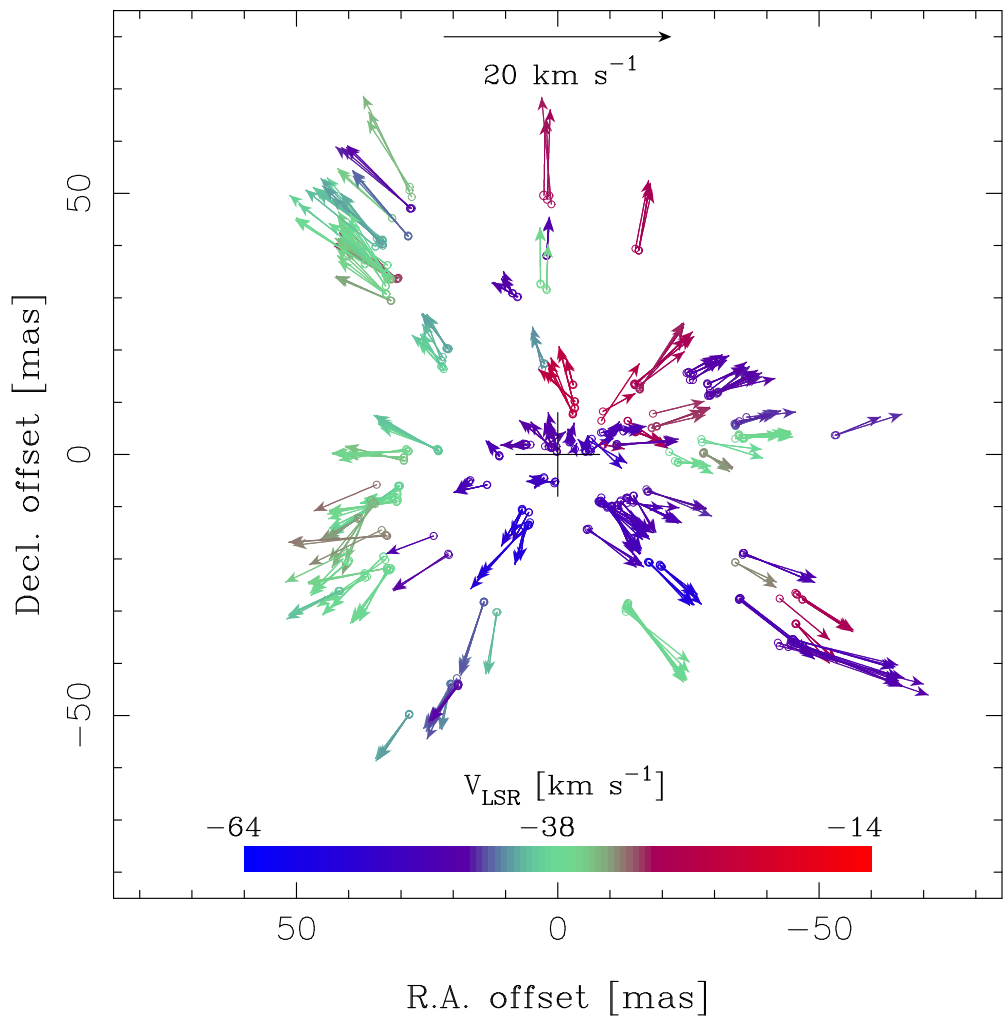}
\caption{
    Internal proper motions of maser spots of S~Per. 
    The color represents the radial velocity. 
}
\label{fig:10}
\end{figure}

The obtained proper motions of the maser spots are considered 
to be a combination of an expanding flow of the circumstellar 
envelop (CSE) and the proper motion of the red supergiant 
in the 
Milky Way. 
We finally conducted a model fitting 
analysis assuming a spherically expanding flow model described by 
\citet{Imai2000} 
and 
\citet{Imai2003}. 
In the model fitting, we estimated the distribution centroid indicating 
the stellar position as well as the stellar proper motion separated 
from the expanding shell motion of the CSE. 
We obtained the stellar proper motion in right ascension as 
$\mu^{*}_{\alpha}\cos{\delta} = -0.49 \pm 0.23$~mas~yr$^{-1}$, 
and in declination as 
$\mu^{*}_{\delta} = -1.19 \pm 0.20$~mas~yr$^{-1}$. 
This proper motion is consistent with that 
listed in the Tycho-2 Hipparcos catalogue, 
($-0.1 \pm 1.8$,~$-2.8 \pm 1.8$)~mas~yr$^{-1}$, 
within its uncertainty 
\citep{Hog2000}. 
The centroid of the maser distribution was determined 
with an accuracy of 8~mas in both right ascension and declination. 
Results of our astrometric analysis for S~Per are listed 
in Table~\ref{tbl:05}. 
A residual time variation of the maser emission after 
removing 
the annual parallax and stellar proper motion is shown in 
Figure~\ref{fig:09}. Internal proper motions 
of the identified 478 maser spots are shown in Figure~\ref{fig:10} 
along with their radial velocities. 

\section{
  Discussions
}\label{sec:05}


\subsection{
  Distance to S~Per
}\label{sec:05-01}

Our monitoring program determined the trigonometric distance 
to S~Per to be 
$2.42^{+0.11}_{-0.09}$~kpc 
without 
model 
assumptions. 
\citet{Marvel1996} 
estimated a distance to this star 
as 
$2.3\pm0.5$~kpc by means of a 
{\it statistical parallax} method, 
in which relative proper motions 
of the H$_2$O maser spots are fitted to an expanding shell 
where proper motions of the outermost spots are equal to the 
radial velocities of the maser spots with no significant proper 
motions. The statistical parallax distance 
coincides with the trigonometric one within the measurement 
uncertainties. 
The trigonometric parallax greatly improved the accuracy of the 
distance: the uncertainty is five times better than that of the 
statistical parallax. 

S~Per is considered to belong to Per~OB1 because the radial velocity 
of S~Per is consistent with that of this OB association 
\citep{Bidelman1947}. 
The double cluster, $h+\chi$~Per, whose distance was determined 
to be $2.34$~kpc with the 2-percent accuracy from the H-R diagram 
\citep{Slesnick2002} 
is also included in Per~OB1, so that S~Per is in the vicinity of 
the double cluster in the OB association. The apparent radius 
of Per~OB1 is about $4^{\circ}$, corresponding to 
a $\sim 170$-pc radius at the distance of $h+\chi$~Per. S~Per is 
located on the far side of Per~OB1. 

Recent VLBI astrometric observations of 
Galactic H$_{2}$O and CH$_{3}$OH maser 
sources in the Perseus arm have been used to measure 
their trigonometric distances. 
Figure~\ref{fig:11} shows the locations of such Galactic masers 
close to S~Per in the Perseus arm in a face-on view of 
the 
Milky Way 
\citep{Xu2006, 
Sato2008, 
Moellenbrock2009, 
Moscadelli2009,
Reid2009a}. 
In the figure, we depict only the H$_{2}$O masers for NGC~281~West; 
measurements using the methanol masers result in 
a smaller trigonometric distance than using the H$_{2}$O masers 
\citep{Rygl2010}. 
From this figure, it is clear that S~Per, as well as NGC~281~West, 
are located at the outer edge of the Perseus arm. 
Assuming a pitch angle of the spiral arm of $16.5^{\circ}$ 
\citep{Reid2009b}, 
we can estimate the width of the Perseus arm from the difference in 
locations between the outer edge maser sources and those located 
in the inner part of the arm. Hence, the estimated 
width of the Perseus arm is at least $\sim 0.7$~kpc. 

\subsection{
  Three-dimensional motion of S~Per in the 
Milky Way 
}\label{sec:05-02}

The stellar proper motion of S~Per enables us to discuss 
the three-dimensional kinematics of the evolved star in 
the 
Milky Way. 
In this subsection, 
we use realistic values of the Solar motion
\footnote{
  $U_{\odot}$ is the velocity component directed to the Galactic center, 
  $V_{\odot}$ is the component toward the circular rotational direction 
  in the Galactic plane, and 
  $W_{\odot}$ is the component perpendicular to the Galactic plane 
  and directed to the Galactic north pole. 
}, 
($U_{\odot}$,~$V_{\odot}$,~$W_{\odot}$)$=$
(7.5,~13.5,~6.8)~km~s$^{-1}$ 
(Francis \& Anderson 2009; 
see also discussions in 
McMillan \& Binney 2010). 
Let us estimate a non-circular motion of S~Per in 
LSR 
at the red supergiant. This star has a rather large non-circular motion 
of 15~km~s$^{-1}$ 
at the Galactocentric distance of 10.4~kpc: 
the non-circular velocity components are 
$3.3 \pm 2.3$~km~s$^{-1}$ 
toward the Galactic center, 
$12.9 \pm 2.9$~km~s$^{-1}$ 
toward the anti-direction of the Galactic rotation, and 
$6.3 \pm 3.1$~km~s$^{-1}$ 
toward the south pole. 

Recent VLBI astrometric observations of Galactic maser sources 
have revealed large non-circular motions of several objects 
in the Perseus arm. Figure~\ref{fig:11} presents the non-circular 
velocity components of S~Per along with those from 
other maser objects in the Perseus arm. 
Although this non-circular motion is not sensitive to $R_0$ and 
$\Theta_0$ with the 10-percent variation, 
we have to admit that, if there is a 10-percent dip in the Galactic 
rotation curve at the Galactocentric distance of 9--10~kpc 
\citep{Sofue2009}, 
the lag of S~Per from the Galactic rotation ($V$ component) 
becomes smaller. However, such a rotation curve cannot remove all 
the observed non-circular motions of the maser sources in the Perseus 
arm, or even makes several of them larger. 
Moreover, such peculiar motions can be found for Galactic masers 
in star forming regions (SFRs) not only along the Perseus arm, 
but also along the local spur, Outer and Inner arms 
\citep{Reid2009b, 
Rygl2010}. 
These two papers also discuss a large lag of the $V$ component 
of the SFRs from a flat rotation curve together with a Hipparcos 
Solar motion of 
\citet{Dehnen1998}. 
On the other hand, 
\citet{McMillan2010} reanalyzed the samples published by 
\cite{Reid2009b} 
in order to estimate the Galactic rotation as well as the Solar 
motion. Their fitting result of the $V_{\odot}$ component of 
11~km~s$^{-1}$, instead of 5.2~km~s$^{-1}$ of 
\citet{Dehnen1998}, 
can reduce such 
the 
large lags. However the Galactic SFRs still seems to 
have $10-20$~km~s$^{-1}$ peculiar motions even with the revised Solar 
motion, and those non-circular velocity components in the Galactic plane 
somehow show randomness. 
These non-circular motions do not quantitatively match a standard 
understanding of the stationary density-wave theory 
\citep{Lin1964} 
and a standing galactic shock solution in a tight-winding spiral 
potential 
\citep{Roberts1969, Roberts1970}.  

One of the reasons for such peculiar motions may be attributed to 
an intrinsic space motion of each SFR. 
\citet{Lee2008} 
found that there are new SFRs in the southern 
$H_{\mathrm{I}}$ shell in Per~OB1, and that the new stars have 
larger southern velocities due to the up-to-date cloud shuffling 
caused by $h+\chi$~Per. 
According to the cloud-shuffling model, a molecular cloud is swept up 
by UV radiation pressure from nearby OB stars, 
perhaps together with supernova explosions. Star formations are 
triggered in the cloud 
to form OB associations. At the same time, the compressed cloud 
will be accelerated 
in the anti-direction to the original sources causing the cloud shuffling. 
\cite{Cappa2000}
reported on an H$_{\mathrm{I}}$ superbubble with $2^{\circ}$-radius 
centered at $h+\chi$~Per. However, S~Per, located inside the superbubble, 
does not move away from the double cluster, so that this star does not 
seem to be formed in a process of the cloud shuffling caused 
by $h+\chi$~Per. 

Although we cannot find a trace of cluster shuffling sources other than 
$h+\chi$~Per at this moment, the cloud shuffling model still provides us 
an attractive idea in order to explain the large non-circular motion of 
the OB association and S~Per. 
It is interesting that 
\citet{Lee2008}
also reported that proper motions of stars in Per~OB1 
based on the 
Hipparcos 
measurements show a velocity component downward to the Galactic plane 
(toward the South Galactic pole), so that the non-circular motion of 
S~Per may be representative of a motion of the molecular cloud in which 
the OB stars and S~Per were formed. The plausible scenario is 
as follows: 
%
%
Considering the age of $h+\chi$~Per is $12.8 \pm 1.0$~Myrs 
\citep{Slesnick2002}, 
a cloud shuffling took place near the precursor molecular cloud 
of Per~OB1 around a few tens of Myrs ago because of radiation pressure 
from precedent OB stars and/or supernova explosions in the Galactic 
plane. The molecular cloud later forming Per~OB1 has been accelerated 
in the out-of-disk direction. 
S~Per was formed in the molecular cloud ahead of a large SFR 
around the double cluster. This cloud shuffling 
also might affect the in-disk non-circular motions of the cloud. 
Another item of observational evidence for a cloud shuffling causing 
the non-circular motion was provided by 
\citet{Sato2008} 
and 
\citet{Rygl2010} for NGC~281~West. It 
was 
suggested that 
large non-circular motions of NGC~281~West masers 
were 
due to 
the supershell generated by supernova explosions. 

Another new idea to explain the non-circular motions of the Galactic SFRs 
arises 
from 
large N-body/hydrodynamics numerical simulations. 
\citet{Baba2009} 
showed that, with their high-resolution numerical model of a spiral 
galaxy and their original N-body/gas simulation code 
\citep{Saitoh2009}, 
the non-circular motions of the SFRs 
were 
explained by natural 
consequence of non-linear interactions between non-stationary spiral 
arms and multi-phase interstellar media. With their numerical simulations 
of a disk galaxy, they found that young stars with ages $\leq 50$~Myr 
in a spiral galaxy have large and random non-circular motions.  
This idea can be applied to S~Per and Per~OB1 because the ages of 
those objects 
are younger than 50~Myr. 
The N-body+hydrodynamics simulations also revealed that molecular 
clouds in a spiral galaxy can have a velocity dispersion of 
several km~s$^{-1}$ in the vertical direction to the disk 
(J.~Baba, 2009, private communications), 
so that a rather large peculiar motion of Per~OB1 vertical to the 
Galactic plane could be explained without a cloud shuffling. 
Since we have quite limited samples with the measurable non-circular 
motions of the young objects in the 
Milky Way, 
it is very important to measure the distances and proper motions 
of the spiral arm objects to compare with those simulation results. 
A deep understanding of the dynamics of the Galactic spiral arms 
and the origin of the non-circular motions of SFRs will be possible 
in near future when massive numerical 
simulations can be combined with a much larger amount of VLBI 
astrometry samples. 

\section{
  Conclusions
}\label{sec:06}

We have conducted phase-referencing VLBI monitoring for the Galactic 
red supergiant, S~Per. 
The distance to S~Per was determined to be 
$2.42^{+0.11}_{-0.09}$~kpc 
based on 
the trigonometric parallax. Together with the stellar radial 
velocity and obtained proper motion, we investigated 
the three-dimensional motion of S~Per in the 
Milky Way, 
showing the non-circular motion of 15~km~s$^{-1}$ 
from the Galactic circular rotation, 
which is mainly dominated by the anti rotation direction component 
of $12.9 \pm 2.9$~km~s$^{-1}$. 
This non-circular motion may be 
representative of the motion of the young star cluster, Per~OB1, 
in the 
Milky Way. 
Recent large N-body/hydrodynamics numerical simulations show that 
the non-circular motions of 
star forming regions 
are not exceptional but can be seen everywhere in a spiral galaxy. 
The cloud shuffling is another plausible mechanism to give such a large 
non-circular motion to the Per~OB1 association. To reveal these 
non-circular motions as well as the other maser sources on the 
spiral arms, a lot of sample points are needed. 
In 
the 
near future, we will be able to obtain a set of trigonometric measurements 
with the three-dimensional motions 
of the maser sources with phase-referencing VLBI.  
We will be able to investigate the galactic structure along with 
its kinematics with a support of the large-scale N-body/hydrodynamics 
numerical simulations. 


\acknowledgments

The authors acknowledge the referee for carefully reading our paper 
and giving us many fruitful comments. 
The authors also express their deep gratitude to the VLBA of the NRAO. 
The VLBA/NRAO is a facility of the National Science Foundation, 
operated under a cooperative agreement by Associated Universities, Inc. 
Y.~Asaki would like to thank J.~Baba, K.~Wada, J.~Makino, 
and T.~Saito of NAOJ for suggestions about the Galactic rotation. 
H.~Imai was financially supported by Grant-in-Aid for Scientific 
Research from Japan Society for Promotion Science (20540234).

\newpage

\clearpage
\onecolumn

\clearpage
%
%
%
%
\clearpage
%
%
\clearpage
%
%
\begin{figure}
\epsscale{.50}
\plotone{fig04.epsi}
\caption{
    Radio spectrum of KR143. The abscissa is an observing frequency, 
    and the ordinate is a flux density (both in logarithmic scales). 
    The open square, 
    the filled square, 
    the open triangle, 
    the open diamond, 
    the filled triangle, and 
    the filled circle 
    represent 
    the Texas survey 
\citep{Douglas1996}, 
    21-cm radio continuum survey 
\citep{Kallas1980}, 
    NRAO VLA Sky Survey (NVSS) 
\citep{Condon1998}, 
    the 6-cm northern sky catalogue 
\citep{Becker1991}, 
    the outer galaxy VLA survey 
\citep{Fich1986}, 
    and the J-net VLBI observations 
\citep{Imai2001}, 
    respectively. The open circles represent the current VLBA 
    observation results which are averaged at each frequency. 
}
\label{fig:04}
\end{figure}
\clearpage
%
%
%
\clearpage
%
%
%
\clearpage
%
%
%
\clearpage
%
%
%
\clearpage
%
%
%
%
\clearpage
%
%
%
\clearpage
%
%
%
\begin{figure}
\epsscale{.55}
\plotone{fig11.epsi}
\caption{
    Spatial positions and non-circular velocity components of S~Per 
    and other maser sources in the Perseus arm projected on the 
    Galactic plane. 
    The depicted sources are 
    W3~(OH) 
    \citep{Xu2006}, 
    NGC~281~West 
    \citep{Sato2008}, 
    IRAS~00420+5530 
    \citep{Moellenbrock2009}, 
    S~252 
    \citep{Reid2009a},
    NGC~7538
    \citep{Moscadelli2009},
    and S~Per (present paper). 
    A dark-gray dashed curve indicates the Perseus arm obtained 
    from H$_{\mathrm{II}}$ survey 
    \citep{Georgelin1976}  
    while a light-gray dashed curve indicates the one obtained 
    from CO survey 
    \citep{Nakanishi2006}. Those traces are adjusted to those 
    for $R_{0}=8.5$~kpc. 
    Dotted curves indicate equi-distances from the 
    Galactic center (8, 10, and 12~kpc). 
}
\label{fig:11}
\end{figure}
\clearpage
%
%


\clearpage
%
%
%
\begin{table}
\begin{center}
\caption{
    Observation epochs and 
    duty-cycle times 
    for VLBI phase-referencing.
}
\label{tbl:01}
\begin{tabular}{cp{30mm}cp{45mm}p{45mm}}
\tableline\tableline
        &    &    & \multicolumn{1}{c}{$t_{\mathrm{s}}$ at 22~GHz}
                  & \multicolumn{1}{c}{$t_{\mathrm{s}}$ at 15~GHz} \\
  Epoch & \multicolumn{1}{c}{Date} 
        & \multicolumn{1}{c}{Time range}
        & \multicolumn{1}{l}{~(KR143 --}
        & \multicolumn{1}{l}{~(KR143 --} \\
        &
        & \multicolumn{1}{c}{(UTC)}
        & \multicolumn{1}{r}{S~Per)}
        & \multicolumn{1}{r}{ICRF~0244+624)}  \\
  \tableline
    A   & 2000 Nov 21   & 03:10 -- 08:10 & \multicolumn{1}{c}{40~s}
                          & \multicolumn{1}{c}{--}    \\
    B   & 2002 May 3    & 16:26 -- 21:26 & \multicolumn{1}{c}{40~s}         
                          & \multicolumn{1}{c}{--}   \\
    C   & 2002 May 29   & 14:44 -- 19:44 & \multicolumn{1}{c}{40~s}
                          & \multicolumn{1}{c}{--}   \\
    D   & 2003 Jul 13   & 10:48 -- 17:48 & \multicolumn{1}{c}{35~s}
                          & \multicolumn{1}{c}{60~s} \\
    E   & 2003 Oct 7    & 05:10 -- 12:09 & \multicolumn{1}{c}{35~s}
                          & \multicolumn{1}{c}{60~s} \\
    F   & 2004 May 2    & 15:29 -- 22:27 & \multicolumn{1}{c}{35~s}
                          & \multicolumn{1}{c}{60~s} \\
    G   & 2004 Aug 11   & 08:46 -- 15:46 & \multicolumn{1}{c}{35~s}
                          & \multicolumn{1}{c}{60~s} \\
    H   & 2007 Feb 22   & 20:03 -- 03:01 & \multicolumn{1}{c}{35~s}
                          & \multicolumn{1}{c}{60~s} \\
\tableline
\end{tabular}
\end{center}
\end{table}
\clearpage
%
%
\begin{table}
\tabletypesize{scriptsize}
\begin{center}
  \caption{Observed pairs of sources in our monitoring program.}
  \label{tbl:02}
  \begin{tabular}{lcccc}
  \tableline\tableline
    Target        & Phase-referencing   & Separation  & Frequency  & Epoch \\
                  & calibrator          & angle       &  \\
  \tableline
    KR143\tablenotemark{a}
                  & S~Per               & $0.2^\circ$  & 22.2~GHz  & A -- H \\
    KR143         & ICRF~0244+624\tablenotemark{b}
                                        & $4.8^\circ$  & 15.3~GHz  & D -- H \\
  \tableline
  \end{tabular}
  \tablenotetext{a}{
        KR143 is a positional reference of S~Per in VLBI phase-referencing 
        at 22~GHz. 
  }
  \tablenotetext{b}{
        ICRF~0244+624 is a positional reference of KR143 in VLBI 
        phase-referencing at 15~GHz. 
  }
  \end{center}
\end{table}
\clearpage
%
%
\begin{table}
  \begin{center}
  \caption{Determined KR143's positions. }
  \label{tbl:03}
  \begin{tabular}{lccc}
\tableline\tableline
    Source   & RA (J2000) & Dec (J2000) \\
\tableline
    KR143\tablenotemark{a}
 & $2^{\mathrm{h}} 22^{\mathrm{m}} 33^{\mathrm{s}}.520836$~$\pm$~0$^s$.000018
 & $+58^{\circ} 48' 13''.944455$~$\pm$~0$''$.000256 \\
\tableline
\end{tabular}
\tablenotetext{a}{
        The positional reference in the International Celestial Reference 
        Frame (ICRF) is ICRF~0244+624. 
        The position error of ICRF~0244+624 in ICRF 
        (0.124 and 0.228~mas in right ascension and declination, 
        respectively) is taken into consideration. 
        This position is located at the origin of 
        Figure~\ref{fig:02} and of the top plot in Figure~\ref{fig:03}. 
}
\end{center}
\end{table}
\clearpage
%
%
\begin{deluxetable}{ccccccc}
\tablewidth{0pt}
\tablecaption{
  Proper motions and parallaxes for the S~Per H$_2$O maser spots. 
  \label{tbl:04}
}
\tablehead{
\colhead{Spot ID           } & 
\colhead{$\Delta\alpha_{A}$} & 
\colhead{$\Delta\delta_{A}$} &
\colhead{V$_{\mathrm{LSR}}$} &
\colhead{$\pi$             } &
\colhead{$\mu_{\alpha}
          \cos{\delta}$    } &
\colhead{$\mu_{\delta}$    } \\
\colhead{(Feature          } &
\colhead{[mas]             } &
\colhead{[mas]             } &
\colhead{[km~s$^{-1}$]     } &
\colhead{[mas]             } &
\colhead{[mas~yr$^{-1}$]   } &
\colhead{[mas~yr$^{-1}$]   } \\
\colhead{ID)               } &
\colhead{                  } &
\colhead{                  } &
\colhead{                  } &
\colhead{                  } &
\colhead{                  } &
\colhead{                  }
}
\startdata
 $1(1)$ & $  -27.85 \pm     0.11$ & $    0.11 \pm     0.03$ & $-33.58$ & $0.417 \pm 0.047$ & $ -0.95 \pm   0.04$ & $ -1.40 \pm   0.01$ \\
 $2(1)$ & $  -27.86 \pm     0.10$ & $    0.26 \pm     0.03$ & $-33.79$ & $0.398 \pm 0.041$ & $ -0.93 \pm   0.03$ & $ -1.43 \pm   0.01$ \\
 $3(1)$ & $  -27.86 \pm     0.10$ & $    0.28 \pm     0.03$ & $-34.00$ & $0.398 \pm 0.040$ & $ -0.93 \pm   0.03$ & $ -1.44 \pm   0.01$ \\
 $4(1)$ & $  -27.89 \pm     0.10$ & $    0.27 \pm     0.03$ & $-34.21$ & $0.425 \pm 0.042$ & $ -0.92 \pm   0.03$ & $ -1.42 \pm   0.01$ \\
 $5(1)$ & $  -27.98 \pm     0.12$ & $    0.48 \pm     0.04$ & $-34.42$ & $0.410 \pm 0.051$ & $ -0.89 \pm   0.04$ & $ -1.50 \pm   0.01$ \\
 $6(2)$ & $  -17.31 \pm     0.17$ & $   -7.16 \pm     0.03$ & $-45.59$ & $0.403 \pm 0.046$ & $ -1.35 \pm   0.05$ & $ -1.43 \pm   0.01$ \\
 $7(3)$ & $  -46.82 \pm     0.10$ & $  -27.82 \pm     0.03$ & $-29.57$ & $0.412 \pm 0.039$ & $ -1.23 \pm   0.03$ & $ -1.66 \pm   0.01$ \\
 $8(3)$ & $  -45.69 \pm     0.12$ & $  -26.81 \pm     0.03$ & $-29.57$ & $0.454 \pm 0.039$ & $ -1.37 \pm   0.04$ & $ -1.73 \pm   0.01$ \\
 $9(3)$ & $  -46.83 \pm     0.10$ & $  -27.79 \pm     0.03$ & $-29.78$ & $0.429 \pm 0.040$ & $ -1.21 \pm   0.03$ & $ -1.66 \pm   0.01$ \\
$10(3)$ & $  -45.84 \pm     0.10$ & $  -26.86 \pm     0.03$ & $-29.78$ & $0.470 \pm 0.038$ & $ -1.32 \pm   0.03$ & $ -1.71 \pm   0.01$ \\
$11(3)$ & $  -45.46 \pm     0.09$ & $  -26.52 \pm     0.03$ & $-30.21$ & $0.352 \pm 0.043$ & $ -1.37 \pm   0.03$ & $ -1.79 \pm   0.01$ \\
$12(4)$ & $  -14.57 \pm     0.15$ & $   -7.89 \pm     0.04$ & $-47.06$ & $0.475 \pm 0.052$ & $ -0.81 \pm   0.05$ & $ -1.76 \pm   0.01$ \\
$13(4)$ & $  -13.52 \pm     0.14$ & $   -8.53 \pm     0.03$ & $-47.69$ & $0.431 \pm 0.041$ & $ -1.21 \pm   0.04$ & $ -1.50 \pm   0.01$ \\
$14(4)$ & $  -13.24 \pm     0.14$ & $   -8.37 \pm     0.03$ & $-48.12$ & $0.411 \pm 0.042$ & $ -1.30 \pm   0.04$ & $ -1.54 \pm   0.01$ \\
$15(4)$ & $  -12.08 \pm     0.11$ & $   -8.78 \pm     0.03$ & $-48.12$ & $0.372 \pm 0.039$ & $ -0.88 \pm   0.04$ & $ -1.72 \pm   0.01$ \\
$16(4)$ & $  -13.28 \pm     0.14$ & $   -8.37 \pm     0.03$ & $-48.33$ & $0.412 \pm 0.043$ & $ -1.29 \pm   0.04$ & $ -1.54 \pm   0.01$ \\
$17(4)$ & $  -11.36 \pm     0.13$ & $  -10.00 \pm     0.03$ & $-48.33$ & $0.449 \pm 0.042$ & $ -0.90 \pm   0.04$ & $ -1.54 \pm   0.01$ \\
$18(4)$ & $   -7.85 \pm     0.12$ & $   -9.12 \pm     0.03$ & $-49.38$ & $0.404 \pm 0.043$ & $ -1.26 \pm   0.04$ & $ -1.72 \pm   0.01$ \\
$19(4)$ & $   -7.81 \pm     0.11$ & $   -8.95 \pm     0.03$ & $-49.59$ & $0.386 \pm 0.043$ & $ -1.27 \pm   0.04$ & $ -1.77 \pm   0.01$ \\
$20(5)$ & $   41.93 \pm     0.10$ & $  -26.14 \pm     0.04$ & $-39.90$ & $0.384 \pm 0.051$ & $  0.28 \pm   0.03$ & $ -1.57 \pm   0.01$ \\
$21(5)$ & $   41.93 \pm     0.10$ & $  -26.15 \pm     0.03$ & $-40.11$ & $0.403 \pm 0.043$ & $  0.27 \pm   0.03$ & $ -1.56 \pm   0.01$ \\
$22(5)$ & $   41.88 \pm     0.10$ & $  -26.15 \pm     0.03$ & $-40.32$ & $0.425 \pm 0.042$ & $  0.27 \pm   0.03$ & $ -1.54 \pm   0.01$ \\
$23(5)$ & $   41.83 \pm     0.10$ & $  -26.15 \pm     0.03$ & $-40.53$ & $0.401 \pm 0.045$ & $  0.28 \pm   0.03$ & $ -1.56 \pm   0.01$ \\
$24(6)$ & $   20.53 \pm     0.10$ & $  -43.94 \pm     0.03$ & $-43.48$ & $0.472 \pm 0.047$ & $ -0.15 \pm   0.03$ & $ -1.91 \pm   0.01$ \\
$25(6)$ & $   20.29 \pm     0.10$ & $  -44.08 \pm     0.03$ & $-43.69$ & $0.332 \pm 0.042$ & $ -0.08 \pm   0.03$ & $ -1.91 \pm   0.01$ \\
$26(6)$ & $   20.04 \pm     0.10$ & $  -44.47 \pm     0.03$ & $-43.90$ & $0.405 \pm 0.041$ & $ -0.03 \pm   0.03$ & $ -1.82 \pm   0.01$ \\
$27(6)$ & $   20.12 \pm     0.12$ & $  -44.67 \pm     0.03$ & $-44.11$ & $0.425 \pm 0.041$ & $ -0.07 \pm   0.04$ & $ -1.77 \pm   0.01$ \\
$28(6)$ & $   20.28 \pm     0.14$ & $  -44.78 \pm     0.03$ & $-44.32$ & $0.419 \pm 0.042$ & $ -0.13 \pm   0.04$ & $ -1.75 \pm   0.01$ \\
$29(6)$ & $   20.27 \pm     0.12$ & $  -44.75 \pm     0.03$ & $-44.53$ & $0.426 \pm 0.041$ & $ -0.13 \pm   0.04$ & $ -1.78 \pm   0.01$ \\
$30(6)$ & $   19.21 \pm     0.10$ & $  -44.09 \pm     0.03$ & $-44.74$ & $0.425 \pm 0.040$ & $ -0.05 \pm   0.03$ & $ -1.72 \pm   0.01$ \\
$31(6)$ & $   20.26 \pm     0.13$ & $  -44.81 \pm     0.03$ & $-44.74$ & $0.418 \pm 0.044$ & $ -0.14 \pm   0.04$ & $ -1.78 \pm   0.01$ \\
$32(6)$ & $   19.10 \pm     0.10$ & $  -43.90 \pm     0.03$ & $-44.96$ & $0.417 \pm 0.039$ & $ -0.04 \pm   0.03$ & $ -1.78 \pm   0.01$ \\
$33(6)$ & $   19.05 \pm     0.10$ & $  -43.97 \pm     0.03$ & $-45.17$ & $0.408 \pm 0.039$ & $ -0.03 \pm   0.03$ & $ -1.75 \pm   0.01$ \\
$34(6)$ & $   19.03 \pm     0.10$ & $  -44.06 \pm     0.03$ & $-45.38$ & $0.419 \pm 0.039$ & $ -0.03 \pm   0.03$ & $ -1.73 \pm   0.01$ \\
$35(6)$ & $   19.03 \pm     0.10$ & $  -44.14 \pm     0.03$ & $-45.59$ & $0.430 \pm 0.039$ & $ -0.04 \pm   0.03$ & $ -1.72 \pm   0.01$ \\
$36(6)$ & $   19.06 \pm     0.10$ & $  -44.22 \pm     0.03$ & $-45.80$ & $0.433 \pm 0.041$ & $ -0.06 \pm   0.03$ & $ -1.73 \pm   0.01$ \\
$37(6)$ & $   19.11 \pm     0.10$ & $  -44.32 \pm     0.04$ & $-46.01$ & $0.454 \pm 0.050$ & $ -0.09 \pm   0.03$ & $ -1.74 \pm   0.01$ \\
$38(7)$ & $   28.43 \pm     0.09$ & $  -49.64 \pm     0.03$ & $-41.79$ & $0.430 \pm 0.037$ & $  0.02 \pm   0.03$ & $ -1.93 \pm   0.01$ \\
$39(7)$ & $   28.47 \pm     0.09$ & $  -49.79 \pm     0.03$ & $-42.01$ & $0.430 \pm 0.037$ & $ -0.00 \pm   0.03$ & $ -1.89 \pm   0.01$ \\
$40(7)$ & $   28.46 \pm     0.09$ & $  -49.80 \pm     0.03$ & $-42.22$ & $0.422 \pm 0.037$ & $ -0.01 \pm   0.03$ & $ -1.89 \pm   0.01$ \\
$41(7)$ & $   28.48 \pm     0.09$ & $  -49.83 \pm     0.03$ & $-42.43$ & $0.428 \pm 0.037$ & $ -0.04 \pm   0.03$ & $ -1.87 \pm   0.01$ \\
$42(8)$ & $   -9.84 \pm     0.15$ & $    3.81 \pm     0.03$ & $-47.91$ & $0.404 \pm 0.041$ & $ -0.67 \pm   0.05$ & $ -1.03 \pm   0.01$ \\
$43(8)$ & $   -9.79 \pm     0.11$ & $    3.91 \pm     0.03$ & $-48.12$ & $0.378 \pm 0.039$ & $ -0.68 \pm   0.03$ & $ -1.06 \pm   0.01$ \\
$44(8)$ & $   -9.80 \pm     0.11$ & $    3.98 \pm     0.03$ & $-48.33$ & $0.351 \pm 0.039$ & $ -0.67 \pm   0.03$ & $ -1.07 \pm   0.01$ \\
\enddata
\end{deluxetable}
\clearpage
%
%
\begin{table}
  \begin{center}
  \caption{Astrometry analysis results of S~Per. }
  \label{tbl:05}
  \begin{tabular}{lcc}
\tableline\tableline
     Annual parallax   
   & \multicolumn{2}{c}{
       $\pi^{*} = 0.413 \pm 0.017$~mas\ \ 
       ($2.42^{+0.11}_{-0.09}$~kpc)} \\
\tableline
    Position (J2000)\tablenotemark{a}
  & Right Ascension: 
  & Declination: \\
    (2000 Nov 21)
  & $2^{\mathrm{h}} 22^{\mathrm{m}} 51^{\mathrm{s}}.7106$~$\pm$~0$^s$.0010
  & $+58^{\circ} 35' 11''.444$~$\pm$~0$''$.008 \\
\tableline
    Stellar proper motion 
  & $\mu^{*}_{\alpha}\cos{\delta}=-0.49 \pm 0.23$~mas~yr$^{-1}$
  & $\mu^{*}_{\delta}=-1.19 \pm 0.20$~mas~yr$^{-1}$ \\
\tableline
  \end{tabular}
\tablenotetext{a}{
        The position errors of ICRF~0244+624 and KR143 
        are taken into consideration as well as the model fitting 
        error described in Section~\ref{sec:04-02}. 
        This position is located at the origin of S~Per image 
        of Figure~\ref{fig:05}.}
\end{center}
\end{table}
\clearpage
%
%
%



\clearpage




\end{document}